\RequirePackage{lineno}

\documentclass[aps,pre,oldlfont,amsmath,amsfonts,amssymb,onecolumn]{revtex4}
\usepackage{bm}        

\usepackage[latin1]{inputenc}
\usepackage{graphicx}
\usepackage[english]{babel}
\usepackage[usenames,dvipsnames]{color}

\usepackage{amsfonts}
\usepackage{amsmath}
\usepackage{amssymb}

\usepackage{color}

\usepackage{soul}

\newcommand{\beq}{\begin{equation}}
\newcommand{\eeq}{\end{equation}}
\newcommand{\nn}{\nonumber}
\newcommand{\ket}[1]{|#1\rangle}
\newcommand{\bra}[1]{\langle #1|}

\newcommand{\ra}{\rightarrow}

\makeatletter
\@ifundefined{textcolor}{}
{%
 \definecolor{BLACK}{gray}{0}
 \definecolor{WHITE}{gray}{1}
 \definecolor{RED}{rgb}{1,0,0}
 \definecolor{GREEN}{rgb}{0,1,0}
 \definecolor{BLUE}{rgb}{0,0,1}
 \definecolor{CYAN}{cmyk}{1,0,0,0}
 \definecolor{MAGENTA}{cmyk}{0,1,0,0}
 \definecolor{YELLOW}{cmyk}{0,0,1,0}
 }

\begin{document}


\title{Exact solution of a lambda quantum system driven by a two-photon wavepacket}

\author{W. Lopes
$^{1}$
}

\author{D. Valente
$^{1}$
}
\email{valente.daniel@gmail.com}

\affiliation{
$^{1}$ 
Instituto de F\'isica, Universidade Federal de Mato Grosso, Cuiab\'a, MT, Brazil
}


\begin{abstract}
Three-level atoms in lambda configuration find diverse applications in quantum information processing, and a promising way to manipulate their quantum states is with single-photon pulses propagating in a waveguide (which can be theoretically regarded as a highly broadband regime of the Jaynes-Cummings model).
Here, we analytically find the non-perturbative dynamics of a lambda atom driven by a two-photon wavepacket, propagating in a one-dimensional electromagnetic environment.
As an application, we study the dynamics of a quantum state purification.
By comparing our exact model with an approximated model of two cascaded single-photon wavepackets, we show how two-photon nonlinearities and stimulated emission affect the purification.
\end{abstract}


\maketitle
\section{Introduction}
Two-level quantum systems can play the role of qbits, i.e., quantum bits of information.
Implementing a logical qbit within the two ground states of a three-level atom in $\Lambda$ configuration has the advantage that the excited state provides an energetic barrier, blocking unwanted jumps between the two target states, thus increasing the stability of physical operations.
This motivates various applications of the $\Lambda$ atoms, such as in quantum memories \cite{rmp2017, molmer}, quantum cloning \cite{dvnjp,dv2012,qsr,csimon1,csimon2}, and quantum state purification \cite{qda, cp22}.

To achieve efficient manipulation of the atom, one can either employ optical cavities or one-dimensional (1D) waveguides.
Waveguides are especially desirable if one wishes that a propagating single-photon wavepacket interacts with a single atom, for instance if the photon works as a flying qubit making distant atoms in a network to communicate
\cite{kimble,lodahl, nori, jmg}.
From a theoretical viewpoint, these engineered 1D environments can be described as the highly broadband regime of the Jaynes-Cummings (JC) model \cite{chen2004,fan,domokos}.
This has unraveled a rich diversity of effects, including non-Markovianity \cite{nonmarkov}, broadband photon-photon correlations \cite{roy,njp17,lodahl22}, irreversible stimulated emission \cite{dvnjp,continuum2018,dv2012}, as well as the application of quantum mirrors in Fabry-Perot \cite{prlfratini} and Mach-Zehnder \cite{nelson} interferometers.

Here, our main motivation is to understand how the type of quantum state purification for a $\Lambda$ atom as proposed in Ref.\cite{qda} is affected by a two-photon wavepacket.
To that end, we present an exact non-perturbative solution, in real-space representation, for the dynamics of a two-photon wavepacket interacting with a single $\Lambda$ system in a broadband one-dimensional environment.
We describe light-matter interaction by employing the broadband JC model, and analytically compute the quantum state of the composite atom-plus-field system within the two-excitation subspace.
We obtain the transition probability between the two ground states of the atom, depending on the initial two-photon pulse parameters.
In order to highlight the presence of two-photon nonlinearities, we compare our exact results with the effective model of two cascaded (consecutive) single-photon wavepackets.

\begin{figure}[!htb]
\centering
\includegraphics[width=0.7\linewidth]{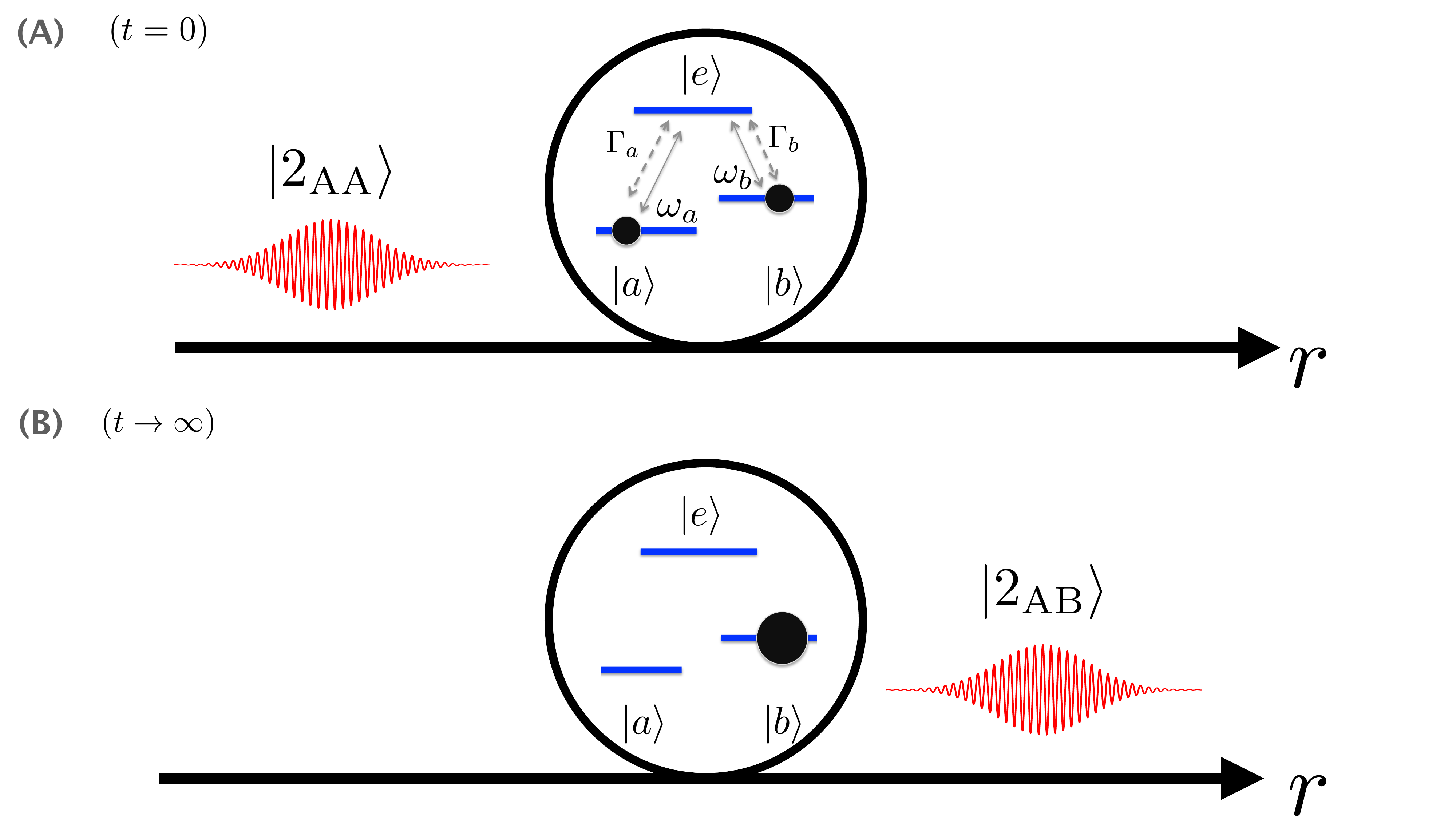} 
\caption{
Quantum state purification with a two-photon wavepacket in a one-dimensional environment with spatial coordinate $r$.
(A) At time $t=0$, a two-photon wavepacket $\ket{2_{AA}}$, with $AA$-polarized photons, both resonant with transition frequency $\omega_a$, starts interacting with the atom.
The photons are described by their linewidths (inversely related with the spatial spread of the pulse).
The atom is initially in a generic mixture of $\ket{a}$ and $\ket{b}$ states.
The atom decay rates are $\Gamma_{a}$ and $\Gamma_{b}$, respectively between states $\ket{e}$ to $\ket{a}$, and $\ket{e}$ to $\ket{b}$.
(B) Asymptotically, $t\ra \infty$, the atom may have been left at the pure state $\ket{b}$, as governed by the transition probability $p_{a \ra b}(\infty)$.
The quantum state of the outgoing wavepacket is altered to $\ket{2_{AB}}$, both in spacetime and in polarization (in this case, to $AB$).
Our broadband JC model, within a real-space representation of the amplitudes, allows us to compute the time-dependent probability amplitudes for the global atom-plus-field system.
}
\label{fig1}
\end{figure}

\section{Model}
We consider a global time-independent Hamiltonian of the system and its environment,
\beq
H=H_S + H_I + H_E.
\label{global}
\eeq
We model light-matter interaction via a dipole coupling, in the rotating-wave approximation \cite{domokos, chen2004, fan, dvnjp, dv2012, nonmarkov},
\beq
H_I = -i\hbar \sum_\omega (g_a a_\omega \sigma_a^\dagger + g_b b_\omega \sigma_b^\dagger - \mbox{H.c.}),
\label{HI}
\eeq
where $\sigma_k = \ket{k}\bra{e}$ (for $k=a,b$) and $\mbox{H.c.}$ is the Hermitian conjugate.
Orthogonal polarization modes $\left\{ a_\omega \right\} $ and $\left\{ b_\omega \right\}$ respectively interact with transitions $\ket{a}$ to $\ket{e}$ and $\ket{b}$ to $\ket{e}$.
This is a multi-mode Jaynes-Cummings model, and our interest is in the limit of a broadband continuum of frequencies, 
\beq
\sum_\omega \ra \int d\omega \varrho_\omega \approx \varrho \int d\omega,
\label{rho}
\eeq
where we have defined a constant density of modes $\varrho_\omega \approx \varrho$.
Physically, this means that there is no frequency filter in our idealized environment.
The continuum of frequencies allows us to employ a Wigner-Weisskopf approximation to obtain dissipation rates
$\Gamma_{a} = 2\pi g_a^2 \varrho$ and 
$\Gamma_{b} = 2\pi g_b^2 \varrho$.
The environment Hamiltonian reads
\beq
H_E = \sum_\omega \hbar \omega (a_\omega^\dagger a_\omega + b_\omega^\dagger b_\omega).
\eeq
The system is described by a three-level Hamiltonian, with states named $\ket{a}$, $\ket{b}$, and $\ket{e}$, so that
\beq
H_S = \hbar \omega_a \ket{e}\bra{e} + \hbar \delta_{ab} \ket{b}\bra{b}.
\eeq
When in $\Lambda$ configuration, these energy levels provide an elementary energy barrier (the excited state $\ket{e}$) separating two highly stable states (the two ground states $\ket{a}$ and $\ket{b}$).
The energy difference between states $\ket{b}$ and $\ket{a}$, given by $\hbar\delta_{ab} = \hbar(\omega_a - \omega_b)$, although relevant for thermal equilibrium considerations, does not affect our results as far as nonequilibrium effects are concerned.

As stated above, our main interest here is in applying our exact solution to the study of quantum state purification.
This is the process where the reduced state of the atom asymptotically goes from a mixed to a pure state, such as
\beq
p_a^{(0)} \ket{a}\bra{a}
+
p_b^{(0)} \ket{b}\bra{b}
\ra
\ket{b}\bra{b},
\eeq
after being driven by the photons and dissipated to the environment.
To achieve this kind of dynamics, we focus on the case where the initial state of the field is a two-photon wavepacket polarized at modes $\left\{ a_\omega \right\}$.
This means that the photons will leave the atom unaltered, if initially at $\ket{b}$.
Thus the reduced state of the atom evolves as
\beq
p_a^{(0)} 
\mbox{tr}_F[ U \ket{a, 2_{AA}}\bra{a, 2_{AA}} U^\dagger ]
+
p_b^{(0)} \mbox{tr}_F[ U \ket{b, 2_{AA}}\bra{b, 2_{AA}} U^\dagger ]
=
p_a^{(0)} \mbox{tr}_F[ U \ket{a, 2_{AA}}\bra{a, 2_{AA}} U^\dagger ]
+
p_b^{(0)}
\ket{b} \bra{b},
\eeq
where $\mbox{tr}_F$ is the partial trace over the field degrees of freedom, 
and $U= \exp(-iHt/\hbar)$.
We thus have to focus on the $\ket{a} \ra \ket{b}$ transition probability, defined as
\beq
p_{a \ra b}(t) \equiv \bra{b} \mbox{tr}_F[ U \ket{a, 2_{AA}}\bra{a, 2_{AA}} U^\dagger ] \ket{b},
\eeq
if we wish to study the purification dynamics of the atomic state.
If $p_{a \ra b}(\infty) = 1$, the atom will be left at the pure state $\ket{b}$, asymptotically.
See Fig.(\ref{fig1}).

\section{Results}

\subsection{Quantum state of the global atom-plus-field system in the two-excitation subspace}
We take advantage of the number conservation as provided by the JC model in order to restrict ourselves to the two-excitation subspace, spanned by
\begin{eqnarray}
    \ket{\psi}
    =\sum_{\omega}\psi_{\omega}^{A}\hat{a}_{\omega}^\dagger \ket{e,0} 
    +\sum_{\omega}\psi_{\omega}^{B}\hat{b}_{\omega}^\dagger \ket{e,0} 
    +\sum_{\omega_{1}\omega_{2}}\phi_{\omega_{1}\omega_{2}}^{AA} \hat{a}_{\omega_{1}}^\dagger \hat{a}_{\omega_{2}}^\dagger \ket{a,0}
    +2 \sum_{\omega_{1}\omega_{2}}\phi_{\omega_{1}\omega_{2}}^{AB} \hat{a}_{\omega_{1}}^\dagger \hat{b}_{\omega_{2}}^\dagger \ket{a,0}  \nn \\
    + \sum_{\omega_{1}\omega_{2}}\phi_{\omega_{1}\omega_{2}}^{BB} \hat{b}_{\omega_{1}}^\dagger \hat{b}_{\omega_{2}}^\dagger \ket{b,0}
    + 2\sum_{\omega_{1}\omega_{2}}\phi_{\omega_{1}\omega_{2}}^{BA} \hat{b}_{\omega_{1}}^\dagger \hat{a}_{\omega_{2}}^\dagger \ket{b,0}.
\label{psi}
\end{eqnarray}

By substituting that into the Schr\"odinger equation, $i\hbar \partial_t \ket{\psi} = H \ket{\psi}$, where $H$ is given in Eq.(\ref{global}), we find that

\beq
\partial_t \psi_{\omega}^{A}(t) = -i(\omega_{a}+\omega)\psi_{\omega}^{A}(t)-2\sum_{\omega'}[g_{a}\phi_{\omega\omega'}^{AA}(t) + g_{b}\phi_{\omega\omega'}^{BA}(t)].
\eeq
\beq
    \partial_t \psi_{\omega}^{B}(t)=-i(\omega_{a}+\omega)\psi_{\omega}^{B}(t)-2\sum_{\omega'}[g_{b}\phi_{\omega\omega'}^{BB}(t) + g_{a}\phi_{\omega\omega'}^{AB}(t)];
\eeq
\beq
    \partial_t\phi_{\omega\omega'}^{AA}(t) = -i(\omega+\omega')\phi_{\omega\omega'}^{AA}(t)+\frac{1}{2}(g_{a}\psi_{\omega}^{A}(t) + g_{a}\psi_{\omega'}^{A}(t))
\eeq
\beq
        \partial_t\phi_{\omega\omega'}^{BB}(t) = -i(\omega+\omega'+\delta_{ab})\phi_{\omega\omega'}^{BB}(t) + \frac{1}{2} (g_{b}\psi_{\omega}^{B}(t) + g_{b}\psi_{\omega'}^{B}(t) );
\eeq
\beq
    \partial_t\phi_{\omega\omega'}^{AB}(t) = -i(\omega+\omega')\phi_{\omega\omega'}^{AB}(t)+\frac{1}{2}g_{a}\psi_{\omega'}^{B}(t);
\eeq
\beq
    \partial_t\phi_{\omega\omega'}^{BA}(t) = -i(\omega+\omega'+\delta_{ab})\phi_{\omega\omega'}^{BA}(t)+\frac{1}{2}g_{b}\psi_{\omega}^{A}(t).
\eeq

\subsection{Formal solution of the amplitudes}

Inspired by the Wigner-Weisskopf theory of the spontaneous emission, we first integrate the field equations,
\begin{eqnarray}
    \phi_{\omega\omega'}^{BA}(t)=\phi_{\omega\omega'}^{BA}(0)e^{-i(\omega+\omega'+\delta_{ab})t} 
    +\frac{1}{2}e^{-i(\omega+\omega'+\delta_{ab})t}\int_{0}^{t}e^{i(\omega+\omega'+\delta_{ab})t'}g_{b}\psi_{\omega}^{A}(t')dt',
    \label{phiBA} 
\end{eqnarray}
and
\begin{eqnarray}
    \phi_{\omega\omega'}^{AA}(t)=\phi_{\omega\omega'}^{AA}(0)e^{-i(\omega+\omega')t}+\frac{1}{2}e^{-i(\omega+\omega')t}\int_{0}^{t}e^{i(\omega+\omega')t'}\left(  g_{a}\psi_{\omega}^{A}(t')+g_{a}\psi_{\omega'}^{A}(t') \right ) dt'.
    \label{phi AA}
\end{eqnarray}

We then substitute them back into the excited-state amplitudes,
\begin{eqnarray}
    \frac{\partial}{\partial t} \psi_{\omega}^{A}(t)&=&-i(\omega_{a}+\omega)\psi_{\omega}^{A}(t)-2\sum_{\omega'} [g_{a} \phi_{\omega\omega'}^{AA}(0)e^{-i(\omega+\omega')t}+ g_{b}\phi_{\omega\omega'}^{BA}(0)e^{-i(\omega+\omega'+\delta_{ab})}  + \\
    \nonumber
    &-& g_{a}\frac{1}{2}\int_{0}^{t} e^{-i(\omega+\omega')(t-t')}g_{a} \psi_{\omega}^{A}(t')dt'  -
    g_{a} \frac{1}{2}\int_{0}^{t} e^{i(\omega+\omega')(t-t')}g_{a}\psi_{\omega'}^{A}(t')dt' + \\
    &-& g_{b}\frac{1}{2}\int_{0}^{t} e^{i(\omega+\omega'+\delta_{ab})(t-t')}g_{b}\psi_{\omega}^{A}(t') dt' ].
\end{eqnarray}
 
Because we are interested in the case where the initial state of the atom is $\ket{a}$ and the field is prepared with two photons at polarization $A$, we have only shown the solutions that are relevant to this scenario.
The more general case can be easily obtained by performing the same steps as used here.

\subsection{Real-space representation}

We define the real-space representations of all the amplitudes, namely,
\begin{eqnarray}
    \psi^{A}(r,t)=\sum_{\omega}\psi_{\omega}^{A}(t)e^{ik_{\omega}r},
    \label{psireal}
\end{eqnarray}

and
\begin{eqnarray}
    \phi^{AA}(r_1,r_2,t)=\sum_{\omega , \omega'}\phi_{\omega,\omega'}^{AA}(t)e^{ik_{\omega} r_1 + ik_{\omega'} r_2},
    \label{phireal}
\end{eqnarray}
with analogous expressions for the $B$ components.
Here, we use a linear approximation for the dispersion relation, $k_\omega = \omega/c$, with $c$ being the speed of light.
The idea that only $k_\omega > 0$ is considered can be justified in two ways: either with unidirectional waveguides employing chiral couplings between atom and field, or considering that each polarization mode is a symmetric superposition of two counter-propagating modes in a bidirectional waveguide.

In these representations, the equations of motion read
\begin{eqnarray}
     \nonumber
     \left[ \frac{\partial}{\partial t}+ c \frac{\partial}{\partial r}\right]\psi^{A}(r,t) = -\left( \frac{\Gamma}{2}+i\omega_{a} \right )\psi^{A}(r,t)-2g_{a}\phi^{AA}(r-ct,-ct,0) \\
     -\Gamma_{a}\psi^A(-r,t-r/c)\Theta(r)\Theta\left( t-\frac{r}{c} \right),
     \label{eqpsiA}
\end{eqnarray}
where we have defined $\Gamma\equiv \Gamma_{a}+\Gamma_{b}$, as obtained after we employ the Wigner-Weisskopf approximation,
\begin{eqnarray}
    \Gamma_{a,b}=\sum_{\omega'}2\pi g_{a,b}^{2}\delta(\omega_{a,b}-\omega') = 2\pi g_{a,b}^2 \varrho.
\end{eqnarray}
Because the coupling $g_a$ is not experimentally accessible in this regime (only $\Gamma_a$ is), we take $g_a = \sqrt{\Gamma_a/(2\pi \varrho)}$, where $\varrho$ is the density of modes defined in Eq.(\ref{rho}) (the density of modes $\varrho$ cancels out at the end, so no probabilities will depend on it).
Another crucial remark concerning the Wigner-Weisskopf approximation employed here is that, at some point in the calculation, we have an integration of the type 
$\int_0^t dt' \delta (t' - \tau)$, 
where $\tau = t-r/c$.
Because $0 < t' < t$, this term amounts to zero, unless $0 < \tau < t$, which gives us $0 < t-r/c < t$;
hence $0 < t-r/c$ justifies the Heaviside step function $\Theta(t-r/c)$, whereas $t-r/c < t$ gives us $r > 0$, so $\Theta(r)$ appears.
In Eq.(\ref{eqpsiA}), we have also set
$\phi_{\omega\omega'}^{BA}(r-ct,-ct,0) = 0$, 
since we are considering that no $B$-polarized photons are incident on the atom.

The interpretation of Eq.(\ref{eqpsiA}) is that of a wave equation propagating towards the positive direction in space.
The damping term accounts for the spontaneous emission of the atom, so the homogeneous solution is 
$\psi^A(r-ct,0) \exp[-(\Gamma+i\omega_a) t]$.
The second term on the right-hand side of Eq.(\ref{eqpsiA}) corresponds to a pair of photons driving the atom, with strength proportional to the coupling frequency $g_a$.
If $g_a = g_b = 0$, the field freely propagates at speed $c$, so its amplitude will obey to
$\phi^{AA}(r_1,r_2,t) = \phi^{AA}(r_1 - ct , r_2 - ct , 0)$, in that case.
This allows us to understand the following reasoning.
The atom is at position $r_a=0$, so if there is a chance that a photon starts at position $r_2 = -ct$, then this photon will reach the atom after a time $t$.
This explains the dependence of the driving amplitude on $-ct$.
The other photon started at $r_1 = r - ct$, explaining why $\phi^{AA}(r-ct,-ct,0)$ also depends on $r-ct$.
Finally, the last term on the right-hand side describes stimulated emission.
To see that, we notice that $\Theta(r)$ is nonzero only for $r>0$, so $\psi^A(-r,t-r/c)$ is the excited-state amplitude evaluated at a negative position, $-r < 0$, where the presence of a photon (with $A$ polarization) can stimulate the emission of the excited atom.

\subsection{Solution of the excited-state amplitude in the real-space representation}

Because Eq.(\ref{eqpsiA}) presents a discontinuity due to the Heaviside step functions, we proceed to solve it by a stepwise partitioning the space into three domains, as follows.

\begin{itemize}

\item Region I: $r < 0$.

    \begin{eqnarray}
      \left [\frac{\partial}{\partial t}+c\frac{\partial}{\partial r} \right] \psi_{1}^{A}(r,t)=- \left ( \frac{\Gamma}{2}+i\omega_{a} \right) \psi_{1}^{A}(r,t)- 2g_a\phi^{AA}(r-ct,-ct,0).
      \label{eqR1}
  \end{eqnarray}

\item Region II: $0 < r < ct$.

    \begin{eqnarray}
     \nonumber
     \left [\frac{\partial}{\partial t}+c\frac{\partial}{\partial r} \right] \psi_{2}^{A}(r,t)=- \left ( \frac{\Gamma}{2}+i\omega_{a} \right) \psi_{2}^{A}(r,t)- \Gamma_{a} \psi_{1}^{A}(-r,t-r/c) \\
     - 2g_{a} \phi^{AA}(r-ct,-ct,0).
     \label{eqR2}
    \end{eqnarray}
    
Note that the dynamics in Region II depends on the solution of Region I.

\item Region III: $r > ct$.

    \begin{eqnarray}
     \left [\frac{\partial}{\partial t}+c\frac{\partial}{\partial r} \right] \psi_{3}^{A}(r,t)=- \left ( \frac{\Gamma}{2}+i\omega_{a} \right) \psi_{3}^{A}(r,t)- 2g_{a} \phi^{AA}(r-ct,-ct,0).
     \label{eqR3}
    \end{eqnarray}

\end{itemize}

\

To solve for Region I, we look for a solution of the type
\begin{eqnarray}
    \psi_{1}^{A}(r,t)=e^{- \left ( \frac{\Gamma}{2}+i\omega_{a} \right) t}\bar{\psi}_{1}^{A}(r,t),
    \label{s1}
\end{eqnarray}
as motivated by the homogeneous solution.
We thus find that
\begin{eqnarray}
    \hat{\theta}\bar{\psi}_{1}^{A}=-2g_{a} e^{ \left ( \frac{\Gamma}{2}+i\omega_{a} \right)t}\phi^{AA}(r-ct,-ct,0),
    \label{optheta1}
\end{eqnarray}
where we have defined
$\hat{\theta}\equiv \partial_t+c \partial_r$,
so the inhomogeneous solution is given by
\begin{eqnarray}
    \bar{\psi}_{1}^{A}(r,t)=-2g_{a} \int_{0}^{t}e^{ \left ( \frac{\Gamma}{2}+i\omega_{a} \right) t'}\phi^{AA}(r-ct, -ct',0)dt'.
    \label{psibarra1s}
\end{eqnarray}
To make sense of this solution, we can see it as describing one photon driving the atom (term $-ct'$), while the other photon is located at position $r<0$ at time $t > 0$.
A formal justification of the solution can be given by assuming a general decomposition such as
$\phi^{AA}(r-ct,-ct,0)=\sum_{nm}C_{nm}R_{n}(r-ct)T_{m}(-ct)$,
where $C_{nm}$ are generic coefficients, with $R_n$ and $T_m$ being generic functions.
By noting that 
$\hat{\theta} R_n(r-ct) = 0$, 
we reobtain the original differential equation.

Region II is far richer in terms of dynamical behavior, since it depends on the solution of Region I.
Again, we first eliminate the homogeneous part by taking the transformation
\beq
    \psi_{2}^{A}(r,t)=e^{- \left( \frac{\Gamma}{2}+i\omega_{a} \right) t}\bar{\psi}_{2}^{A}(r,t),
\eeq
which implies that
\begin{eqnarray}
    \hat{\theta}\bar{\psi}_{2}^{A}(r,t)=-e^{\left ( \frac{\Gamma}{2}+i\omega_{a} \right)t} \left[ \Gamma_{a}  \psi_{1}^{A}\left( -r,t-r/c \right) + 2 g_{a} \phi^{AA}(r-ct,-ct,0) \right],
    \label{sumsolpsi2barra}
\end{eqnarray}
where, from Eq.(\ref{psibarra1s}), we have that
\begin{eqnarray}
    \psi_{1}^{A}(-r,t-r/c)&=&-2g_{a}e^{-\left ( \frac{\Gamma}{2}+i\omega_{a} \right)(t-r/c)}\int_{0}^{t-r/c} e^{\left ( \frac{\Gamma}{2}+i\omega_{a} \right)t'} \phi^{AA}(-ct,-ct',0)dt'. \\
    &\equiv& F(t,t-r/c).
\label{psiminusr}
\end{eqnarray}
The purpose of defining $F(t,t-r/c)$ above is just to highlight that the solution only depends on $t$ and $t-r/c$.
Because both terms on the right-hand side of Eq.(\ref{sumsolpsi2barra}) are functions of $t$ and $t-r/c$, we can search for a solution in a similar form as we did in Region I, namely,
\begin{eqnarray}
    \bar{\psi}_{2}^{A}(r,t)=&-& 
    \Gamma_{a} \int_{t-r/c}^{t}e^{\left ( \frac{\Gamma}{2}+i\omega_{a} \right)t'}F(t',t-r/c)dt' -2g_{a}\int_{0}^{t}e^{\left ( \frac{\Gamma}{2}+i\omega_{a} \right)t'}\phi^{AA}(r-ct,-ct',0)dt'.
\label{F}
\end{eqnarray}
A crucial remark: the integration in the first term of Eq.(\ref{F}) is starting at $t-r/c$.
Replacing $t-r/c$ by $0$ in that term also provides a correct solution for Eq.(\ref{sumsolpsi2barra}).
However, our final interest is in solving Eq.(\ref{eqpsiA}), to which this makes a key difference.
We further explain why by the end of this section.

Region III has precisely the same equation of motion of Region I, so the solution is
\begin{eqnarray}
    \bar{\psi}_{3}^{A}(r,t)=-2 g_a e^{-\left ( \frac{\Gamma}{2}+i\omega_{a} \right)t} \int_{0}^{t}e^{\left ( \frac{\Gamma}{2}+i\omega_{a} \right)t'} \phi^{AA}(r-ct,-ct',0) dt'.
\end{eqnarray}
If one is interested in considering a two-photon wavepacket that initially occupies only the $r<0$ region (Region I), as we are here, one can assume that
$\phi^{AA}(r_1,r_2,0) = 0$ for $r_1>0$, which amounts to say that
$\phi^{AA}(r-ct,-ct',0) = 0$ for $r>ct$ (that defines Region III).

Putting all three regions together, and adding the homogeneous solution, we get that
\begin{eqnarray}
    \nonumber
    \psi^{A}(r,t)&=&\psi^{A}(r-ct,0)e^{-\left ( \frac{\Gamma}{2}+i\omega_{a} \right)t}- 2 g_{a} e^{-\left ( \frac{\Gamma}{2}+i\omega_{a} \right)t} \int_{0}^{t}dt' e^{\left ( \frac{\Gamma}{2}+i\omega_{a} \right)t'}\phi^{AA}(r-ct,-ct',0)+ \\
    \nonumber
    &+&2\Gamma_{a} g_{a} \Theta(r)\Theta(t-r/c)e^{-\left ( \frac{\Gamma}{2}+i\omega_{a} \right)t}e^{-\left ( \frac{\Gamma}{2}+i\omega_{a} \right)(t-r/c)} \\ 
    &\times& \int_{t-r/c}^{t}dt'e^{\left ( \frac{\Gamma}{2}+i\omega_{a} \right)t'}\int_{0}^{t-r/c}dt''e^{\left ( \frac{\Gamma}{2}+i\omega_{a} \right)t''}\phi^{AA}(-ct',-ct'',0).
    \label{solgeral}
\end{eqnarray}
%
%
%
%
The first term corresponds to two simultaneous events, namely, the free evolution of the single-photon packet, and the spontaneous emission of the atom.
The second term corresponds to one photon exciting the atom, while the other photon freely propagates.
And the third term describes both photons driving the atom, since $\phi^{AA}(-ct, -ct, 0)$ indicates both photons having arrived at the atomic position $r_a = 0$ after time $t$ has passed.

We can now give a justification for the rather unusual integration of Eq.(\ref{F}) (from $t-r/c$ to $t$).
Starting from Eq.(\ref{solgeral}) and substituting it back into Eq.(\ref{eqpsiA}), we get a spurious extra term coming from 
$\partial_r \Theta(r) = \delta(r)$.
The lower limit defined as $t-r/c$ exactly cancels out that spurious term, since
$\delta(r) \int_{t-r/c}^t (...) = 0$.

\subsection{Solution of the field amplitudes in the real-space representation}

Our main interest is in finding an expression for 
$\phi^{BA}(r_1,r_2,t) = \sum_{\omega, \omega'} \phi^{BA}_{\omega, \omega'}(t) \exp(ik_{\omega} r_1 + ik_{\omega'} r_2)$.
Analogous to Eq.(\ref{phiBA}), and choosing the initial condition $ \phi^{BA}_{\omega, \omega'}(0) = 0$, we have that
\begin{eqnarray}
    \nonumber
    \phi^{BA}(r_{1},r_{2},t)&=& \sum_{\omega\omega'}\left( \frac{1}{2} e^{-i(\omega+\omega'+\delta_{ab})t}\int_{0}^{t} e^{i(\omega+\omega'+\delta_{ab})t'}  g_{b}\psi_{\omega}^{A}(t')dt' \right) e^{ik_{\omega}r_{1}+ik_{\omega'}r_{2}} \\
    \nonumber
    &=& \sum_{\omega\omega'}\left( \frac{1}{2} \int_{0}^{t} e^{-i(\omega+\omega'+\delta_{ab})(t-t')}  g_{b}\psi_{\omega}^{A}(t')dt' \right) e^{ik_{\omega}r_{1}+ik_{\omega'}r_{2}} \\
    \nonumber
    &=& \sum_{\omega \omega'}\frac{1}{2}\int_{0}^{t}e^{ik_{\omega}[r_{1}-c(t-t')]}e^{i\frac{\omega'}{c}[r_{2}-c(t-t')]}e^{-i\delta_{ab}(t-t')}g_{b}\psi_{\omega}^{A}(t')dt'\\
    \nonumber
    &=& \frac{1}{2}g_{b}\int_{0}^{t}\left( \sum_{\omega}\psi_{\omega}^{A}(t')e^{ik_{\omega}[r_{1}-c(t-t')]} \right) \sum_{\omega'}e^{i\frac{\omega'}{c}[r_{2}-c(t-t')]}e^{-i\delta_{ab}(t-t')}dt'\\
    \nonumber
    &=& \frac{1}{2}g_{b}\int_{0}^{t}\psi^{A}(r_{1}-c(t-t'),t')\sum_{\omega'}e^{-i\omega'[(t- r_{2} / c )-t']}e^{-i\delta_{ab}(t-t')}dt'.
\end{eqnarray}
We take the continuum limit of the sum and perform a Wigner-Weisskopf approximation once again.
We also have to make sure that $0 \leq t' \leq t$ is consistent with such approximation.
These two inequalities are guaranteed if we multiply the solution by
$\Theta(r_2) \Theta(t-r_2/c)$.
We finally find that
\begin{eqnarray}
    \phi^{BA}(r_{1},r_{2},t)=\frac{\sqrt{2\pi\varrho\Gamma_{b}}}{2} \Theta(r_{2})\Theta\left(t-\frac{r_{2}}{c} \right)
    e^{-i\delta_{ab}r_{2}/{c}} \psi^{A}(r_{1}-r_{2},t-r_{2}/c).
    \label{phiBArealgeral}
\end{eqnarray}

\subsection{Transition probabilities in the real-space representation}

The transition probability between the two ground states of the $\Lambda$ atom, i.e., from $\ket{a}$ to $\ket{b}$ is given by
\beq
p_{a \ra b}(t) = \bra{b} \mbox{tr}_F( \ket{\psi(t)} \bra{\psi(t)} ) \ket{b},
\eeq
provided that $\ket{\psi(0)} = \ket{a, 2_{AA}} = \sum_{\omega_{1}\omega_{2}}\phi_{\omega_{1}\omega_{2}}^{AA}(0) \hat{a}_{\omega_{1}}^\dagger \hat{a}_{\omega_{2}}^\dagger \ket{a,0}$.
Also, $\ket{\psi(t)} = \exp(-iHt/\hbar) \ket{\psi(0)}$,
where $H$ is given in Eq.(\ref{global}),
and $tr_F$ is the partial trace over the field degrees of freedom.

Within the two-excitation subspace, Eq.(\ref{psi}), we have that
\beq
p_{a \ra b}(t) 
= 
\sum_{\omega_{1}\omega_{2}} |\phi_{\omega_{1}\omega_{2}}^{BB}(t)|^{2}
+
4\sum_{\omega_{1}\omega_{2}}|\phi_{\omega_{1}\omega_{2}}^{BA}(t)|^{2}.
\eeq

Here, we are mainly interested in the case where 
$\phi^{BB}_{\omega_1, \omega_2}(0) = 0$, 
implying that 
$\phi^{BB}_{\omega_1, \omega_2}(t) = 0$.
To compute the non-vanishing term, we have to transform it to the real-space representation, finding that
\beq
\sum_{\omega_{1}\omega_{2}}|\phi_{\omega_{1}\omega_{2}}^{BA}(t)|^{2}
=
\frac{1}{(2\pi \varrho c)^{2}} \int_{-\infty}^{\infty}\int_{-\infty}^{\infty}|\phi^{BA}(r_{1},r_{2},t) \big|^{2}dr_{1}dr_{2}.
\eeq
By substituting Eq.(\ref{phiBArealgeral}) in the double integration above, and using Eq.(\ref{solgeral}) for the excited-state amplitude, we have the exact solution of the transition probability for a generic initial pulse $\phi^{AA}(r_1,r_2,0)$.

\subsection{Analytically solvable example}

Let us compute
\begin{align}
p_{a \ra b}(t) 
&= \frac{4}{(2\pi \varrho c)^{2}} \int_{-\infty}^{\infty} dr_2 \int_{-\infty}^{\infty} dr_1\ |\phi^{BA}(r_{1},r_{2},t) \big|^{2} \\
&= \frac{2 \pi \varrho \Gamma_b}{(2\pi \varrho c)^{2}} \int_{0}^{ct} dr_2 \int_{-\infty}^{\infty} dr_1\ |\psi^{A}(r_{1}-r_{2},t-r_2/c) \big|^{2},
\end{align}
for the case where the atom is initially at ground-state $\ket{a}$, so that $\psi^A(r,0) = 0$, and the initial photon pair is prepared by means of the spontaneous emission of two hypothetical uncorrelated atoms, arbitrarily far away from our $\Lambda$ atom, both polarized at $A$.
We can thus choose
\beq
\phi^{AA}(r_1,r_2,0) = N\bigg[ f_1(r_1) f_2(r_2) + f_1(r_2) f_2(r_1) \bigg],
\label{phiaa0}
\eeq
where
\beq
f_k(r) = \Theta(-r) \exp[(\Delta_k/2 + i\omega_{Lk} )r/c],
\label{exp}
\eeq
and $N$ is a normalization constant satisfying $\bra{\psi(0)} \psi(0) \rangle = 1$, 
equivalent to
$2 \sum_{\omega_1, \omega_2} |\phi^{AA}_{\omega_1, \omega_2} (0)|^2 = 1$, 
or in the real-space representation,
\beq
\frac{2}{(2\pi \varrho c)^{2}} \int_{-\infty}^{\infty}\int_{-\infty}^{\infty}|\phi^{AA}(r_{1},r_{2},0) \big|^{2}dr_{1}dr_{2} = 1.
\eeq
Using Eqs.(\ref{phiaa0}) and (\ref{exp}), and also assuming resonant photons, $\omega_{L1} = \omega_{L2}$, we find that
\beq
N = \pi \varrho \sqrt{\frac{\Delta_1 \Delta_2}{1 + r_{12}}},
\eeq
where we have defined the ratio
$r_{12} \equiv 4 \Delta_1 \Delta_2 / (\Delta_1 + \Delta_2)^2$.
Note that $r_{12} \leq 1$, the equality holding if $\Delta_1 = \Delta_2$.

We now have two terms to solve in Eq.(\ref{solgeral}) for $\psi^A(r,t)$.

The first, where the integration of $\phi^{AA}(r-ct,-ct',0)$ appears, amounts to
\beq
\psi^{A1}(r,t)
=
-2Ng_{a}\Theta(t-r/c) e^{-\frac{\Gamma}{2}t} \bigg[ e^{-\frac{\Delta_{1}}{2}(t-\frac{r}{c})} \frac{2}{\Gamma-\Delta_{2}} \left( e^{\left( \frac{\Gamma-\Delta_{2}}{2} \right)t} - 1 \right) + e^{-\frac{\Delta_{2}}{2}(t-\frac{r}{c})} \frac{2}{\Gamma-\Delta_{1}} \left( e^{\left( \frac{\Gamma-\Delta_{1}}{2} \right)t} - 1 \right) \bigg].
\label{psia1}
\eeq
When we compute $\psi^{A1}(r_1-r_2,t-r_2/c)$, we get that $\Theta(t-r_1/c)$ alters the integration from 
$\int_{0}^{ct} dr_2 \int_{-\infty}^{\infty} dr_1$ 
to 
$\int_{0}^{ct} dr_2 \int_{-\infty}^{ct} dr_1$.

The second term, where a double integration of $\phi^{AA}(-ct',-ct'',0)$ appears, amounts to
\begin{eqnarray}
    \psi^{A2}(r,t)&=&\frac{8N}{(\Gamma-\Delta_{1})(\Gamma-\Delta_{2})}g_{a}\Gamma_{a}\Theta(r)\Theta(t-r/c) e^{-\frac{\Gamma}{2}t} e^{-\frac{\Gamma}{2}\left(t-\frac{r}{c} \right)} \nn \\
    &\times& \bigg[ \left( e^{\left ( \frac{\Gamma-\Delta_{1}}{2} \right)t} - e^{\left ( \frac{\Gamma-\Delta_{1}}{2} \right)\left (t-\frac{r}{c} \right)} \right) \left( e^{\left ( \frac{\Gamma-\Delta_{2}}{2} \right)\left (t-\frac{r}{c} \right)} - 1 \right) + \nn \\
    &+& \left( e^{\left ( \frac{\Gamma-\Delta_{2}}{2} \right)t} - e^{\left ( \frac{\Gamma-\Delta_{2}}{2} \right)\left (t-\frac{r}{c} \right)} \right) \left( e^{\left ( \frac{\Gamma-\Delta_{1}}{2} \right)\left(t- \frac{r}{c}\right)} - 1 \right)\bigg].
\label{psia2}
\end{eqnarray}
In the terms proportional to $\psi^{A2}(r_1-r_2,t-r_2/c)$, we have step functions in the form
$\Theta(r_1 - r_2) \Theta(t-r_1/c)$, altering the integration from
$\int_{0}^{ct} dr_2 \int_{-\infty}^{\infty} dr_1$ 
to
$\int_{0}^{ct} dr_2 \int_{r_2}^{ct} dr_1$.

We now have to compute, with the help of Eqs.(\ref{psia1}) and (\ref{psia2}),
\begin{align}
p_{a \ra b}(t) 
&= 
\frac{2 \pi \varrho \Gamma_b}{(2\pi \varrho c)^{2}} \int_{0}^{ct} dr_2 \int_{-\infty}^{ct} dr_1 \big|\psi^{A1}(r_{1}-r_{2},t-r_2/c) \big|^{2} \nn \\
&+
\frac{2 \pi \varrho \Gamma_b}{(2\pi \varrho c)^{2}} \int_{0}^{ct} dr_2 \int_{r_2}^{ct} dr_1 \big|\psi^{A2}(r_{1}-r_{2},t-r_2/c) \big|^{2} \nn \\
&+
\frac{2 \pi \varrho \Gamma_b}{(2\pi \varrho c)^{2}} \int_{0}^{ct} dr_2 \int_{r_2}^{ct} dr_1 
2 
\psi^{A1}(r_{1}-r_{2},t-r_2/c)
\psi^{A2}(r_{1}-r_{2},t-r_2/c),
\end{align}
where the Heaviside step functions have been already considered in the integration limits.

\begin{figure}[!htb]
\centering
\includegraphics[width=1.0\linewidth]{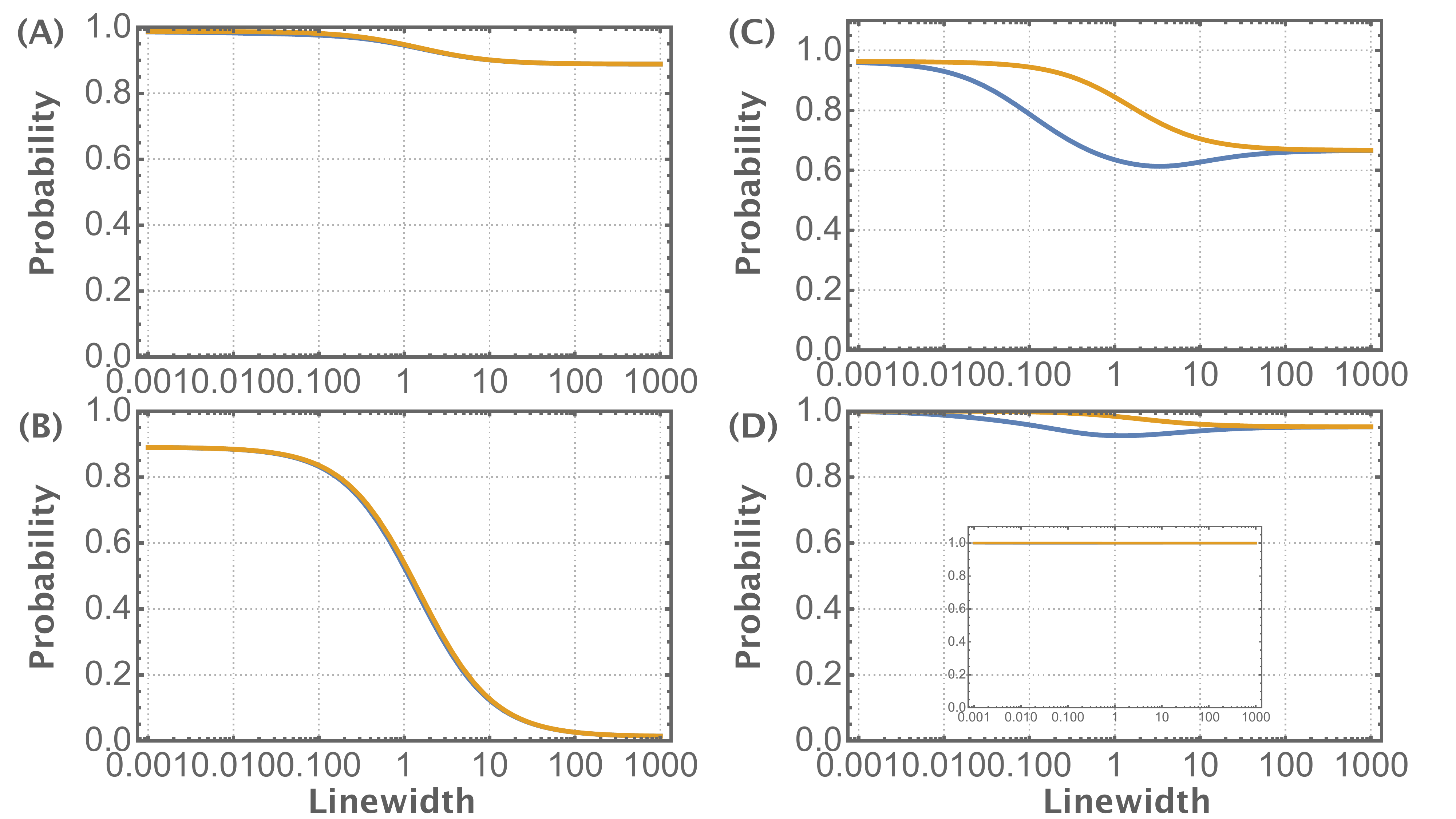} 
\caption{
Asymptotic transition probabilities as functions of the linewidth $\Delta_2$ (in $\Gamma_a$ units).
Exact model: $p_{a\ra b}(\infty)$ (blue).
Approximated (cascaded) model: $p^{(C)}_{a\ra b}(\infty)$ (orange).
We set $\Gamma_a = 1$ in all plots.
(A) $\Gamma_b = 0.5$, and $\Delta_1 = 0.001$.
(B) $\Gamma_b = 0.5$, and $\Delta_1 = 100$.
(C) $\Gamma_b = 0.5$, and $\Delta_1 = 0.5$.
(D) $\Gamma_b=\Gamma_a=1$, and $\Delta_1=0.1$.
Inset: $\Gamma_b=\Gamma_a=1$, and $\Delta_1 = 0.001$.
(A) and (B) show the equivalences between the models, in the very low excitation regimes.
(C) and (D) show the deviations between the models, due to two-photon nonlinearities.
The dips (non-monotonic behaviors) in the blue curves of panels (C) and (D) indicate the presence of stimulated emission.
}
\label{fig2}
\end{figure}

Because the analytical expressions are too cumbersome, in Fig.(\ref{fig2}) we plot
$p_{a \ra b}(\infty)$ 
as a function of $\Delta_2$ (in log-linear scale), for various $\Delta_1$ (blue curves).
We consider both $\Gamma_b = \Gamma_a$ and $\Gamma_b \neq \Gamma_a$ scenarios.

For the sake of a better understanding, we compare our exact two-photon wavepacket model with the approximated model where two cascaded (consecutive, uncorrelated, distinguishable) photons drive the atom (orange curves).
In this cascaded model, we have that the probability transition is given by \cite{ganascini}
\beq
p^{(C)}_{a \ra b}(\infty) = p_1 + (1-p_1)p_2,
\label{pc}
\eeq
with
$p_k = r_{\Gamma}/(1+\Delta_k /\Gamma)$,
and $r_{\Gamma} \equiv 4 \Gamma_a \Gamma_b / \Gamma^2$.
Eq.(\ref{pc}) has a simple interpretation: either the first photon drives the transition, with probability $p_1$, or it leaves the atom unaltered, with probability $1-p_1$, and the second photon drives the transition, with probability $p_2$.
Note that 
$p^{(C)}_{a \ra b}(\infty) = 1$ can only be obtained if $r_{\Gamma} = 1$, which means that equal decay rates, $\Gamma_a = \Gamma_b$, are a necessary condition for a full purification to be achieved with two photons.

Fig.(\ref{fig2})(A) and (B) show regimes where the exact ($p_{a \ra b}(\infty)$, blue) and the approximate ($p^{(C)}_{a \ra b}(\infty)$, orange) models are equivalent (blue and orange curves coincide).
This happens either for too small, $\Delta_1/\Gamma_a = 0.001$ at (A), or too large linewidths, $\Delta_1/\Gamma_a = 100$ at (B).
In both cases, the photons have very small chance of exciting the atom, so two-photon nonlinearities are negligible.

Fig.(\ref{fig2})(C) and (D), by contrast, show how two-photon nonlinearities can violate the cascaded-photons approximation for $\Delta_2/\Gamma_a \sim 1$.
In (C), we have set $\Gamma_b = \Gamma_a/2$, and $\Delta_1 = \Gamma_a / 2$, thus finding a very pronounced deviation.
In (D), we have set $\Gamma_b = \Gamma_a$, and $\Delta_1/\Gamma_a = 0.1$, showing a less pronounced deviation, due to the fact that equal decay rates allow for much closer-to-one transition probabilities then the unequal decay rates scenario.
To give further evidence of this, we also plot the case where 
$\Gamma_a = \Gamma_b$, 
and $\Delta_1 = 0.001$, 
to show that both curves saturate the transition probabilities,
$p_{a \ra b}(\infty) \approx p^{(C)}_{a \ra b}(\infty) \ra 1$ for all values of $\Delta_2$ (inset).

It is important to emphasize the qualitative difference between the two models, as shown in panels (C) and (D): the blue curve is non-monotonic, unlike the orange curve.
This behavior means that overlapping photons of intermediate linewidths ($\Delta_1 \sim \Delta_2 \sim \Gamma$) can suppress (rather than simply not contributing to) the transition from $\ket{a}$ to $\ket{b}$.
In other words, very large linewidths are unable to drive the transition, but intermediate linewidths actively tend to preclude it to take place.
This negative two-photon nonlinearity effect can be understood in terms of a stimulated emission (see discussions in Ref.\cite{dvnjp}, for instance).
If the first photon has appreciable chance of exciting the atom, the presence of another photon of similar linewidth can induce stimulated emission to take place, forcing the atom to go back to initial state $\ket{a}$, instead of transitioning to $\ket{b}$.
This creates the dip in the blue curves of panels (C) and (D).

\section{Conclusions}
We have developed a real-space representation of a two-photon wavepacket driving a single three-level atom in $\Lambda$ configuration.
We have solved the non-perturbative dynamics, as modeled by the broadband JC model.
We have found analytical expressions both for the excited-state amplitudes (see Eq.(\ref{solgeral})) and the ground state amplitudes (see Eq.(\ref{phiBArealgeral})).
Because we were interested in calculating the probability $p_{a \ra b}(t)$ of transition from $\ket{a}$ to $\ket{b}$, we have focused on $\phi^{BA}$, as well as on $\psi^A$.
Nevertheless, our method is directly applicable to compute all the other amplitudes as defined within the two-excitation subspace.
We have also worked out an analytically solvable example, consisting of an initial wavepacket of two photons of exponential envelope shape, resonant with the atom $\ket{a}$ to $\ket{e}$ transition, and analyzed how the probability transition behaves for diverse linewidths of the photons.
We have found that two-photon nonlinearities (taking place when both photons have linewidths comparable to the atomic decay rates) can slightly suppress the transition, due to the non-negligible effect of stimulated emission.

In future studies, we intend to explore further applications of our general solutions.
For instance, we would like to find how a two-photon wavepacket affects quantum cloning processes, such as that from Ref.\cite{qsr}.
Also, we plan to better understand the thermodynamics of this processes. 
How to define work in this case, and how does the work relate with the probability transition from $\ket{a}$ to $\ket{b}$?
These would mean a generalization of the quantum dissipative adaptation as proposed in Ref.\cite{qda}.

\begin{acknowledgements}
This work was supported by CNPq (402074/2023-8) and INCT-IQ (465469/2014-0), Brazil.
W.L. was supported by CAPES.
\end{acknowledgements}


%

%

%

%

%
\end{document}